\documentclass[fleqn,usenatbib]{mnras}

\usepackage{newtxtext,newtxmath}
\usepackage[T1]{fontenc}

\DeclareRobustCommand{\VAN}[3]{#2}
\let\VANthebibliography\thebibliography
\def\thebibliography{\DeclareRobustCommand{\VAN}[3]{##3}\VANthebibliography}

\usepackage{graphicx}	
\usepackage{amsmath}	
\usepackage{multirow}

\title[SIDM and mass stripping of cluster galaxies]{The effects of self-interacting dark matter\\on the stripping of galaxies that fall into clusters}

\author[E.\ L.\ Sirks et al.]{
Ellen L.\ Sirks,\!\thanks{E-mail: ellen.l.sirks@durham.ac.uk}
Kyle A.\ Oman,
Andrew Robertson,
Richard Massey
\& Carlos Frenk
\\
Institute for Computational Cosmology, Department of Physics, Durham University, South Road, Durham DH1 3LE, UK}

\date{Accepted XXX. Received YYY; in original form ZZZ}

\pubyear{2020}

\begin{document}
\label{firstpage}
\pagerange{\pageref{firstpage}--\pageref{lastpage}}
\maketitle

\begin{abstract}
We use the Cluster-EAGLE ({\sc c-eagle}) hydrodynamical simulations to investigate the effects of self-interacting dark matter (SIDM) on galaxies as they fall into clusters. We find that SIDM galaxies follow similar orbits to their Cold Dark Matter (CDM) counterparts, but end up with ${\sim}$25~per~cent less mass by the present day. One in three SIDM galaxies are entirely disrupted, compared to one in five CDM galaxies. However, the excess stripping will be harder to observe than suggested by previous DM-only simulations because the most stripped galaxies form cores and also lose stars: the most discriminating objects become unobservable. The best test will be to measure the stellar-to-halo mass relation (SHMR) 
for galaxies with stellar mass $10^{10-11}\,\mathrm{M}_{\odot}$. This is 8~times higher in a cluster than in the field for a CDM universe, but 13~times higher for an SIDM universe. Given intrinsic scatter in the SHMR, these models could be distinguished with noise-free galaxy-galaxy strong lensing of ${\sim}32$ cluster galaxies. 
\end{abstract}

\begin{keywords}
cosmology: theory -- dark matter -- galaxies: clusters: general -- galaxies: haloes
\end{keywords}

\section{Introduction}
As galaxies fall into clusters, they are transformed, morphologically
and spectroscopically. Their gas content, hitting the intra-cluster
gas, is shocked. Turbulence causes a sudden, final burst of star
formation --- before ram pressure and gravitational tides strip it
away, quenching star formation thereafter
\citep[e.g.][]{Mccarthy_08,2008MNRAS.387...79V,2021MNRAS.501.5073O}. The galaxies'
dark matter (DM) is also eventually stripped by tidal gravity and
gradually becomes incorporated into the (now slightly larger) cluster. This is
the key mechanism for the growth of structure in the Universe; yet, the
timescale for dark matter stripping and virialisation remains poorly
understood. 

In the standard $\Lambda$CDM model of cosmology, dark matter particles
interact with each other only through gravity. The model successfully
explains all observables at large scales, such as the galaxy
clustering signal \citep[for a review see][]{Frenk_White_12} and the the cosmic
microwave background anisotropy
\citep[e.g.][]{2016A&A...594A..13P}. However, there is no a~priori
reason why DM particles should not interact with each other
\citep{2000PhRvL..84.3760S,2000ApJ...534L.143B}, and weak self-interactions
are a natural consequence of some particle physics theories for the
origin of DM \citep[for a review see,
e.g.,][]{2018PhR...730....1T}. With a mean free path ranging from
1\,kpc to 1\,Mpc, DM self-interactions would preserve the large scale
success of $\Lambda$CDM, and could resolve tensions between the
results of DM-only simulations and observations of dwarf and low-mass
galaxies \citep[for a review see][]{2017ARA&A..55..343B}.

Massive galaxy clusters are a promising environment to search for
DM-DM interactions, because the interaction rate would be proportional
to the local DM density and to the local velocity dispersion of DM
particles \citep[for a review see][]{review}. Observations have placed several limits on the strength of
the SIDM cross-section per unit mass ($\sigma/m$) at the typical
velocities encountered in clusters, including
$\sigma/m\lesssim1\,\mathrm{cm}^{2}\,\mathrm{g}^{-1}$ \citep[][from
cluster halo shapes]{2013MNRAS.430..105P}, 
$\sigma/m < 1\,\mathrm{cm}^{2}\,\mathrm{g}^{-1}$ \citep[][from cluster
core sizes]{Rocha_2013}, and
$\sigma/m < 0.47\,\mathrm{cm}^{2}\,\mathrm{g}^{-1}$
(\citealt[][from DM-galaxy offsets in merging
clusters]{2015Sci...347.1462H}, but see also
\citealt{Wittman_2018}). Merging clusters are sufficiently rare that
interpretation of them tends to be limited by uncertainty in their
orientation with respect to the line-of-sight
\citep{2006ApJ...648L.109C,2008ApJ...687..959B,Dawson_2012}. However,
the promising prospects revealed by
\citet{2017MNRAS.465..569R}'s detailed study of high-velocity DM collisions motivates a search for more ubiquitous
examples of objects falling into clusters.

Whenever a galaxy falls into an SIDM cluster, interactions between its
DM particles and those in the cluster could scatter DM out of the
galaxy. This `evaporation' acts in addition to tidal stripping, and
accelerates overall mass loss.  The orbits of infalling galaxies might
also be changed.  Galaxies spiral toward the centre of a cluster due
to dynamical friction, which has strength proportional to the galaxy's
mass \citep[][chapter~8]{2008gady.book.....B}. If galaxies lose
additional mass, they might sink less far or more slowly into the
cluster. On the other hand, drag due to the DM self-interactions
\citep[which may be positive or negative][]{2017MNRAS.465..569R} could
increase the rate of decay; or inhibit the formation of trailing
density wakes in the first place \citep{2017MNRAS.469.2845D}.


The aims of this paper are to study the differences in DM mass loss and orbital dynamics of cluster galaxies, using hydrodynamical simulations with CDM and SIDM physics --- and to investigate whether the differences would be observable. The only previous study of such effects used DM-only simulations \citep{2021arXiv210608292B}.

This paper is organised as follows: in Section~\ref{sec:sim_data}, we
present the simulation suite used in this work; in
Section~\ref{sec:indiv_subs} we study the effects of self-interactions
by matching galaxies between our CDM and SIDM simulations; and in
Section~\ref{sec:observables} we investigate the effects on
observables using the population of galaxies at ${z=0}$. Finally, we
discuss our results and present our conclusions in
Section~\ref{sec:conclusions}.

\section{Data}\label{sec:sim_data}
\subsection{The EAGLE and Cluster-EAGLE simulations}\label{sec:c-eagle}
We use the 50\,Mpc EAGLE cosmological simulation \citep{2015MNRAS.446..521S} and the Cluster-EAGLE ({\sc c-eagle}) zoom cosmological simulations of smaller volumes centred on $\gtrsim10^{14}M_{\odot}$ galaxy clusters \citep{2017MNRAS.470.4186B}.
Both were run with a modified version of the GADGET-3 code that includes radiative cooling, star formation, chemical evolution, and stellar and AGN feedback (with the `AGNdT9' feedback model \citealt{2015MNRAS.446..521S}; \citealt{2015MNRAS.450.1937C}). The DM particle mass is $9.7\times10^{6}\,\mathrm{M}_{\odot}$, the initial gas particle mass is $1.8\times10^{6}\,\mathrm{M}_{\odot}$, and the gravitational softening length was set to 2.66 comoving\,kpc before $z = 2.8$, and then kept fixed at 0.7\,physical\,kpc at $z < 2.8$. The simulations assume cosmological parameters from \cite{2014A&A...571A..16P}. 


The {\sc eagle} volume and two of the {\sc c-eagle} clusters, CE-05 
and CE-12, 
have been re-simulated from identical initial conditions in a $\Lambda$SIDM universe (see Table~\ref{tab:clust_details} and \citealt{2018MNRAS.476L..20R} for more details). These two particular {\sc c-eagle} clusters are `relaxed', based on their gas properties at $z=0.1$ \citep{10.1093/mnras/stx1647}. Since CE-12 is slightly more massive, and has more member galaxies, we shall quote the higher signal-to-noise statistics from that cluster whenever we study the differences between CDM and SIDM at ${z=0}$. However, no data are available for that cluster at higher redshift, so we shall use CE-05 whenever we trace the evolution of DM through time. Note that the central galaxy in CE-05 happened to form early, and the central density cusp has been retained in both CDM and SIDM. The central galaxy of CE-12 formed later, and SIDM interactions created a $\sim100$\,kpc constant density core by ${z=0}$. 



Our implementation of SIDM assumes an isotropic, velocity-independent interaction cross section,  $\sigma/m = 1\,\mathrm{cm}^{2}\,\mathrm{g}^{-1}$. This is around the upper limit of values compatible with current measurements, and therefore maximises the observable consequences. During each simulation timestep, $\Delta t$, DM particles scatter elastically off neighbours within radius $h_\mathrm{SI}=2.66\,\mathrm{kpc}$ (comoving) with probability
\begin{equation}\label{eq:scatter_prob}
P_\mathrm{scat}= \frac{(\sigma/m) ~ m_\mathrm{DM} \, v \, \Delta t}{\frac{4\pi}{3} h_\mathrm{SI}^{3}}~,
\end{equation}
where $v$ is the particles' relative velocity and $m_{\mathrm{DM}}$ the DM particle mass \citep[for more details see][]{2017MNRAS.467.4719R}. We log the time and particle IDs of all DM scattering events. This enables us to distinguish between: DM particles that have not scattered; those that have scattered with other DM particles from their own (sub)halo; and those that have scattered with DM particles from elsewhere in the cluster.

\subsection{Finding and tracking individual galaxies}\label{sec:finding}
We detect groups of particles in the simulations using a {\sc Friends-of-Friends} \citep[FoF, ][]{1985ApJ...292..371D} algorithm with linking length 0.2, and identify individual subhaloes (in all 30 simulation snapshots from $z=14$ to ${z=0}$) using the {\sc subfind} \citep{2001MNRAS.328..726S,2009MNRAS.399..497D} algorithm. For {\sc subfind} to identify a galaxy it must have at least 20 particles. We track subhaloes between snapshots, and construct their merger trees using the {\sc D-Trees} algorithm \citep{2014MNRAS.440.2115J}. 
This identifies each subhalo's $N_\mathrm{link}$ most bound particles of any species, with $N_\mathrm{link}=\mathrm{min}(100,\mathrm{max}(0.1N_\mathrm{gal},10))$, where $N_\mathrm{gal}$ is the total number of particles in the subhalo in each snapshot. The descendant of a subhalo is the object that contains most of these $N_\mathrm{link}$ particles in the next snapshot. A subhalo can have multiple progenitors in the previous snapshot, but we define the main progenitor as that for which the mass summed across all earlier snapshots is the largest. The main branch of a subhalo is comprised of its main progenitors and descendants. We use the main branches of subhaloes to trace their properties through time.

We identify as `field galaxies' all {\sc subfind} central halos (rank 0 in a given FoF group) in {\sc eagle} that contain at least one star particle.
We identify as `cluster member galaxies' all {\sc subfind} subhaloes in {\sc c-eagle} that contain at least one star particle and are within radius $2R_{200}$. We define their time of infall as the first snapshot after they enter that radius for the first time.

The mass of every galaxy is defined as  the total mass, $M_{\mathrm{tot}}$, of all particles gravitationally bound to it (i.e.\ the mass $M_\textsc{sub}$ assigned to the subhalo by the {\sc subfind} algorithm). Its stellar mass, $M_{\star}$, is defined as the total mass of stars within twice its half light radius. Its location is defined by the location of its constituent particle with the lowest gravitational potential energy.

\begin{table}
\centering
\caption{Properties of the CDM and SIDM versions of the two {\sc c-eagle} clusters at redshift ${z=0}$. The mass $M_{200}$ is that enclosed within the sphere of physical radius $R_{200}$ whose mean density is 200 times the critical density of the Universe. Cluster member galaxies are the $N_\mathrm{gal}$ subhalos in the FoF group of the cluster that are within 2$R_{200}$ of the cluster centre and contain one or more star particles.}
\label{tab:clust_details}
\begin{tabular*}{0.88\columnwidth}{@{}c@{\hspace*{15pt}}c@{\hspace*{15pt}}c@{\hspace*{15pt}}c@{\hspace*{15pt}}c@{\hspace*{15pt}}}
\hline
Simulation & DM Type & $M_{200}/\mathrm{M}_{\odot}$ &$R_{200}/\mathrm{Mpc}$ & $N_\mathrm{tot}$ \\
\hline
\multirow{2}{*}{CE-05} & CDM & $1.38\times10^{14}$ & 1.09 & 1442\\
 ~ & SIDM & $1.36\times10^{14}$ & 1.09 & 1183\\
\multirow{2}{*}{CE-12} & CDM & $3.96\times10^{14}$ & 1.55 & 3893\\
~ & SIDM & $3.91\times10^{14}$ & 1.54 & 2938\\
\hline
\end{tabular*}
\end{table}

\begin{figure*}
\centering
\includegraphics[width=0.9\textwidth]{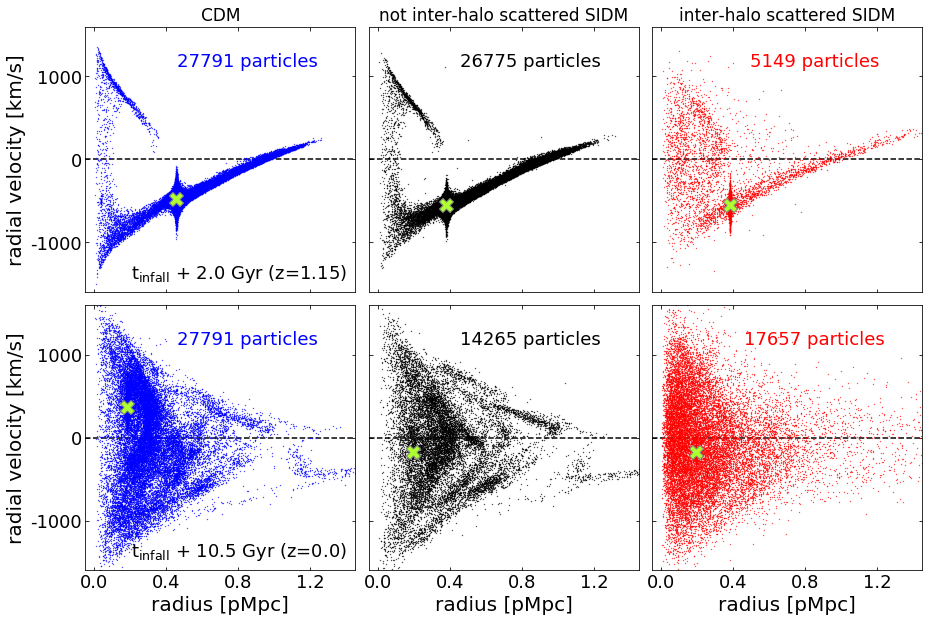}
\caption{Cluster-centric radial velocity as a function of distance from the cluster centre for the DM in a CDM satellite of CE-05 and its SIDM counterpart, 2\,Gyr (top) and 10.5\,Gyr (bottom) after infall. Particles moving outwards from the centre of potential of the cluster have positive radial velocity. Plotted here is the DM that was in the satellite at infall. {\it Left column}: the phase space properties of the DM in the CDM galaxy. {\it Middle column}: the properties of the DM in the SIDM galaxy that has not scattered with the cluster halo DM in the time since infall. {\it Right column}: same as middle column, but for the SIDM that has scattered with the cluster halo since infall. The location of the galaxy itself is indicated by a green cross on each panel.}
\label{fig:sidm_phase_space}
\end{figure*}

\subsection{The stellar-to-halo mass relation}\label{sec:SHMR}

Below we will compare the stellar-to-halo mass relations (SHMR) of galaxies in our SIDM and CDM models. We fit the SHMR to a population of simulated galaxies using the form of the \citet{2013MNRAS.428.3121M} relation derived from abundance matching,
\begin{equation}\label{eq:moster}
M_{\star}(M_{\mathrm{tot}}) =  2NM_{\mathrm{tot}}\left[\left(\frac{M_{\mathrm{tot}}}{M_{1}}\right)^{-\beta}+ \left(\frac{M_{\mathrm{tot}}}{M_{1}}\right)^{\gamma}\right]^{-1}.
\end{equation}
By numerically inverting equation~\eqref{eq:moster}, we also fit $M_{\mathrm{tot}}(M_{\star})$, which can be measured observationally.  

We use the Markov Chain Monte Carlo (MCMC) sampler {\sc emcee} \citep{2013PASP..125..306F} to obtain the best-fit values and posterior PDFs of the free parameters, $M_{1}$, $N$, $\beta$, $\gamma$, as well as the free parameter, $\sigma_{\mathrm{M}}$, the scatter in stellar mass (or in total mass for the inverse fit), which we assume to be constant. The latter enters the fit through the log likelihood function,
\begin{equation}\label{eq:likelihood}
\mathrm{log}\mathcal{L} = -\frac{1}{2}\sum^{N}_{i=1} \left(\frac{\mathrm{log}M_{i} - \mathrm{log}M_{i}^{\mathrm{mod}}}{\sigma_{\mathrm{M}}}\right)^{2}-\frac{N}{2}\mathrm{log}\,\left(2 \pi\sigma_{\mathrm{M}}^{2}\right),
\end{equation}
where the summation is over the total number of galaxies, $N$; $M_{i}$ is the stellar/total mass of galaxy  $i$, and $M_{i}^{\mathrm{mod}}$ is the modelled stellar/total mass of  galaxy $i$, for a given set of parameters. When fitting the SHMR, we truncate fits at the mass where each galaxy includes at least 10 star particles. 

\subsection{Matching galaxies between simulations}\label{sec:matching}

We match galaxies between the CDM and SIDM simulations, so their evolution can be individually compared. In the snapshot after each galaxy crosses $2R_{200}$, we identify its counterpart in the other simulation as that which contains the highest fraction, $f_{\mathrm{match}}$, of shared particle IDs
\begin{equation}\label{eq:matching}
f_{\mathrm{match}} = \frac{N_\mathrm{shared}^{2}}{N_\mathrm{CDM, tot}N_\mathrm{SIDM, tot}},
\end{equation}
where $N_{\rm{shared}}$ is the number of DM particles the CDM galaxy and a possible matching SIDM galaxy have in common, $N_{\rm{CDM, TOT}}$ the total number of DM particles in the CDM galaxy, and $N_{\rm{SIDM, TOT}}$ the total number of DM particles in the SIDM galaxy. 
To complete an association, we require a bijective match: i.e.\ the CDM galaxy points to an SIDM galaxy that points back to it.
The paired CDM and SIDM galaxies inevitably have slightly different infall masses and infall times. When we bin by these, we use the CDM values. This is an arbitrary choice, but none of our results change qualitatively when using either SIDM or common bins (with logarithmic bins of 1\,dex in mass, only 10~per~cent of galaxies are binned differently).

When analysing matched galaxies, we ignore any cluster galaxies that were unmatched to cluster galaxies, and any field galaxies that were unmatched to a central galaxy. In cluster CE-12, 96 out of 889 CDM cluster galaxies were matched to the central galaxy of the SIDM cluster. For the field galaxies, 383 out of 9126 CDM galaxies were matched to a satellite galaxy in the SIDM simulation.

\begin{figure*}
\centering
\includegraphics[width=\linewidth]{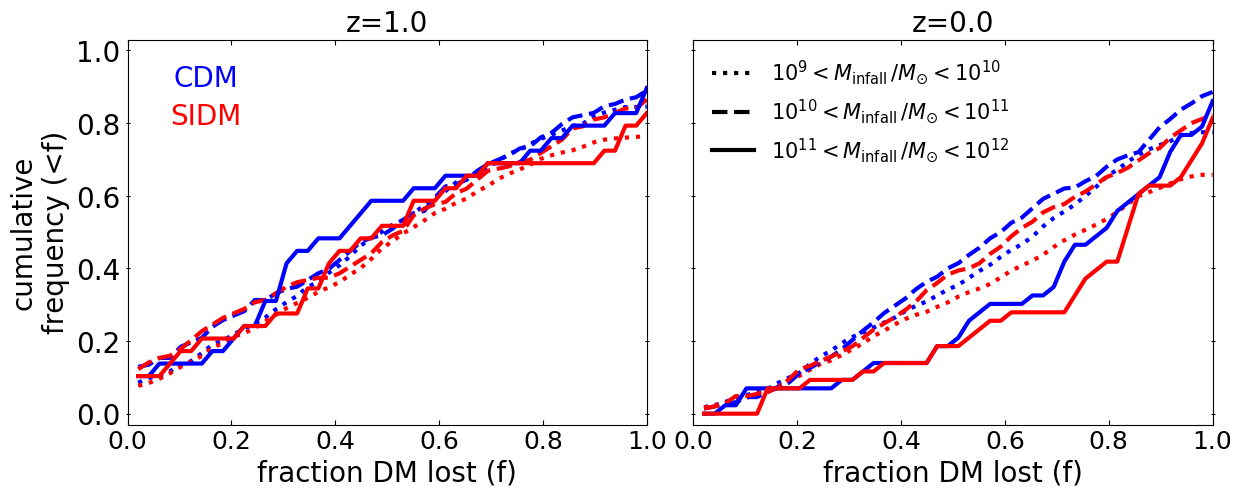}
\caption{The cumulative frequency of galaxies with a fraction of DM
  lost smaller than $f$. Plotted in blue are the distributions for the
  CDM galaxies of CE-05, and in red their SIDM counterparts. The left
  panel shows the results at ${z=1}$, the right panel the results at
  ${z=0}$. The different line styles represent different bins of mass
  at infall, as shown in the legend. A galaxy that has been completely
  disrupted, i.e.\ merged with another galaxy or with the main cluster
  halo is assigned $f=1$. The fraction of disrupted galaxies in each
  infall mass bin is given by 1 minus the cumulative frequency at
  $f=1$, as the cumulative frequency is plotted for fractions smaller
  than $f$.}
\label{fig:phase_space_loss_all}
\end{figure*}

\section{Evolution of DM since infall}\label{sec:indiv_subs}

In this section we examine the effect of self-interactions on the DM
mass of galaxies after they fall into the clusters, by directly
matching galaxies between the CDM and SIDM simulations.


\subsection{The behaviour of one example galaxy}\label{sec:phase_space}
To build intuition, we first track the DM halo of one galaxy in
detail. We identify a typical galaxy that fell into the cluster CE-05
at $z=1.99$ with mass $2.7\times10^{11}\mathrm{M}_{\odot}$, and track
the 6D phase space coordinates (cluster-centric radius and velocity)
of all its DM particles to 2\,Gyr ($z=1.15$) and 10.5\,Gyr (${z=0}$)
after infall. The result is illustrated in
Fig.~\ref{fig:sidm_phase_space} which shows that self-interactions
increase the mass loss of the SIDM galaxy compared to its CDM counterpart, 
but the orbit is unaffected.

\subsubsection{Dark matter loss}

The velocity dispersion of DM within the galaxy is reflected in the `Fingers of God' extending towards positive and negative radial velocities. Tidally stripped DM extends both forwards and backwards along the galaxy's orbit: by 2\,Gyr, some particles have already passed through pericentre and are now moving back out. On a phase-space diagram, tidally stripped material moves along the same path as the galaxy it has been removed from, both in the case of CDM and SIDM. However, the evaporated material should occupy a region distinct from the galaxy and tidally stripped material. 

We separate the SIDM into particles that have scattered with DM particles in the cluster, and particles that have not (Fig.~\ref{fig:sidm_phase_space}). Note that some scattering events result in very low exchange of momentum or merely swap particle trajectories, so the scattered particles include some that have barely been perturbed. However, we find many DM particles that do not follow the tidally stripped material and therefore must be evaporated DM. After 2\,Gyr the CDM galaxy has lost roughly 54~per~cent of its DM mass since infall, whereas the SIDM galaxy has lost approximately 76~per~cent of its DM mass. By ${z=0}$, these fractions have increased to 91~per~cent and 99~per~cent. Evaporation has increased the mass loss in the SIDM galaxy with respect to its CDM counterpart.

We find a much greater SIDM mass loss from galaxies in clusters, than \citet[][fig.~9]{2016MNRAS.461..710D} found for dwarf galaxies in the Milky Way (with the same SIDM cross-section, only a few per~cent more than CDM, 10\,Gyr after accretion). This striking difference is probably due to the much greater DM density and scattering rate in a cluster, but occurs despite the deeper potential wells.

\subsubsection{Orbital evolution}

After 2\,Gyr, the CDM galaxy has moved to a 3D cluster-centric radius of ${\sim}0.4$\,physical~Mpc (pMpc), with a mean radial velocity centred on about $-500\,\mathrm{km}\,\mathrm{s}^{-1}$, i.e.\ the galaxy is moving towards the centre of potential of the cluster (green cross on the top row of Fig.~\ref{fig:sidm_phase_space}). Its SIDM counterpart is within ${\sim}0.1$\,pMpc and has a similar mean radial velocity. By ${z=0}$ the CDM galaxy has moved to a radius of ${\sim}0.2$\,pMpc, with a mean velocity of about $+500\,\mathrm{km}\,\mathrm{s}^{-1}$. Its SIDM counterpart has a mean velocity of about $-100\,\mathrm{km}\,\mathrm{s}^{-1}$, but is located at about the same radius (green cross on the bottom row of Fig.~\ref{fig:sidm_phase_space}). Indeed, we find that there is virtually no difference between the evolution of the 3D cluster-centric radius over time of the CDM and SIDM galaxy (not shown). Self-interactions increase the mass loss of the galaxy, but do not have a significant effect on its orbit.

\subsection{The behaviour of a population of galaxies}\label{sec:pop_subs}

The galaxy used to produce Fig.~\ref{fig:sidm_phase_space} is just one example of the many member galaxies of cluster CE-05. In this section, we investigate the effect of self-interactions on the evolution of DM particles for a large sample of infalling galaxies. 

\subsubsection{Dark matter loss}

In Fig.~\ref{fig:phase_space_loss_all} we plot the cumulative distribution at redshifts\footnote{To be precise, there is no simulation snapshot at exactly ${z=1}$. The snapshot used here is actually at ${z=1.02}$.} ${z=1}$ and ${z=0}$ of the fraction of DM lost from all CDM and SIDM galaxies that were within their cluster in one or more of the 30 simulation snapshots of CE-05. We separate the galaxies into logarithmic bins of 1\,dex in infall mass from $10^{9}$ to $10^{12}\mathrm{M}_{\odot}$. When a galaxy merges with the cluster central galaxy or into the main branch of some other galaxy, we consider it to have been completely disrupted and we set the fraction of DM lost to 1.

At both redshifts, we find that for a given fraction of DM lost, $f$, a greater fraction of the SIDM galaxies have lost a greater portion of their DM than $f$ compared with the CDM galaxies, reflecting the increased mass loss due to self-interactions. The biggest difference is between the low infall mass CDM and SIDM galaxies (dotted lines on Fig.~\ref{fig:phase_space_loss_all}). By ${z=0}$, the mass loss and the difference between the SIDM and CDM galaxies have increased relative to ${z=1}$. We find that a larger fraction of SIDM than CDM cluster galaxies have been disrupted across all mass bins and at both redshifts; see Table~\ref{tab:disrupt}. This is in line with our expectations, as increased mass loss from self-interactions should lead to more disrupted cluster galaxies. 

\begin{table}
\centering
\caption{Fraction of disrupted cluster member galaxies of the CDM and SIDM version of CE-05, at ${z=1}$ and ${z=0}$ and separated into bins of 1\,dex in mass at infall.}
\label{tab:disrupt}
\begin{tabular*}{0.7\columnwidth}{@{}c@{\hspace*{10pt}}c@{\hspace*{10pt}}c@{\hspace*{10pt}}c@{\hspace*{10pt}}c@{}}
\hline
& \multicolumn{2}{c}{$z=1$} & \multicolumn{2}{c}{$z=0$}\\
\hline
Mass range & \multicolumn{2}{c}{$N_{\mathrm{disrupted}}/N_{\mathrm{tot}}$} &  \multicolumn{2}{c}{$N_{\mathrm{disrupted}}/N_{\mathrm{tot}}$}\\
$M_{\odot}$ & \multicolumn{1}{c}{CDM} & SIDM & \multicolumn{1}{c}{CDM} & SIDM\\
\hline
$10^{9}-10^{10}$ & \multicolumn{1}{c}{0.15} & 0.24 & \multicolumn{1}{c}{0.23} & 0.34\\
$10^{10}-10^{11}$ & \multicolumn{1}{c}{0.11} & 0.13 & \multicolumn{1}{c}{0.11} & 0.18\\
$10^{11}-10^{12}$ & \multicolumn{1}{c}{0.1} & 0.17 & \multicolumn{1}{c}{0.14} & 0.19\\
$10^{9}-10^{12}$ & \multicolumn{1}{c}{0.15} & 0.22 & \multicolumn{1}{c}{0.2} & 0.31 \\
\hline
\end{tabular*}
\end{table}

The high mass galaxies (solid lines) have lost a greater fraction of their DM than the galaxies in the other infall mass bins (the solid lines have a different shape than the dotted and dashed lines). This is most likely a consequence of the high mass galaxies having sunk further into the cluster, where stripping becomes more efficient. The timescale for dynamical friction scales with the inverse of the velocity dispersion of the galaxy cubed (section~8.1.1 in \citealt{2008gady.book.....B}), i.e. the time scale decreases as the (infall) mass of the cluster galaxy increases. 

For {\sc subfind} to identify a galaxy it needs to have at least 20 particles. As a consequence a $10^{8}\mathrm{M}_{\odot}$ galaxy can only lose approximately 90~per~cent of its mass before it is already considered disrupted, compared to approximately 99.9~per~cent for a $10^{11}\mathrm{M}_{\odot}$ galaxy. As a result, relatively fewer high mass galaxies disrupt compared to low mass galaxies, even though the high mass galaxies tend to lose a larger fraction of their DM overall. 

The cluster galaxy used to produce Fig.~\ref{fig:sidm_phase_space} has an infall mass of $2.7\times10^{11}\,\mathrm{M}_{\odot}$, placing it in the high mass bin of Fig.~\ref{fig:phase_space_loss_all}. By ${z=0}$, the CDM and SIDM version of this galaxy have lost approximately 91~per~cent and 99~per~cent of their DM mass at infall, corresponding to cumulative frequencies of approximately 0.7 and 0.8 respectively. While both have lost more of their DM than most galaxies of their (high) mass, the loss is not remarkable. 

\subsubsection{Orbital evolution}

\begin{figure}
    \centering
    \includegraphics[width=\linewidth]{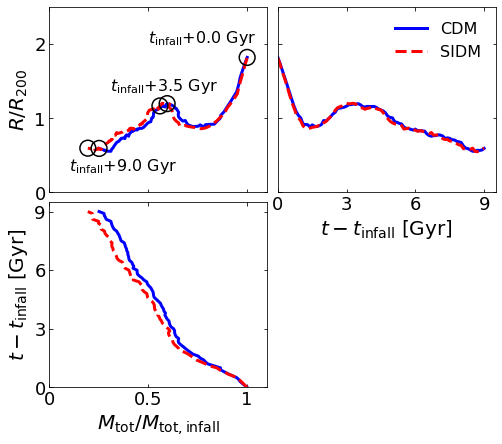}
    \caption{The median evolution since infall of cluster member galaxies with $M_{\star}\gtrsim10^{7}\,\mathrm{M}_{\odot}$ at ${z=0}$, in the CDM (solid blue) and SIDM (dashed red) versions  of cluster CE-05. {\it Top left}: cluster-centric distance in units of $R_{200}$ {\em versus} galaxy mass in units of galaxy mass at infall. The labels indicate the time passed since infall, and the corresponding points on both tracks are encircled. {\it Top right}: cluster centric distance in units of $R_{200}$ as a function of time since infall. {\it Bottom right}: time since infall as a function of galaxy mass in units of the galaxy mass at infall. Note that a different number of galaxies contribute to the median at every point on the plot.}
    \label{fig:stripping_vs_radius}
\end{figure}

We found that the CDM galaxy and its SIDM counterpart used to produce Fig.~\ref{fig:sidm_phase_space} followed nearly the same orbit. To determine whether galaxy orbits in general are unaffected by self-interactions, we now consider the median evolution since time of infall for a sample of galaxies orbiting in the cluster CE-05, in Fig.~\ref{fig:stripping_vs_radius}. We use a sample of 396 matched cluster member galaxies (see Section~\ref{sec:matching}) from CE-05 that have $M_{\star}\gtrsim10^{7}\,\mathrm{M}_{\odot}$ at ${z=0}$. Depending on their infall redshift, the galaxies have spent a different amount of time in the cluster, so a different number of galaxies contribute to each point of Fig.~\ref{fig:stripping_vs_radius}. 

SIDM galaxies start losing more mass than their CDM counterparts about 2\,Gyr after infall (bottom left panel of Fig.~\ref{fig:stripping_vs_radius}). By 9\,Gyr after infall, CDM galaxies have lost $\sim75$~per~cent of their mass, while SIDM galaxies have lost $\sim80$~per~cent. However, we find no difference between the typical orbits of CDM and SIDM galaxies that survive to $z=0$ (top right panel of Fig.~\ref{fig:stripping_vs_radius}; we shall later see very slight differences in the distribution of galaxies that do {not} survive). 

Results are indistinguishable (but noisier) for galaxies with $M_{\star}\gtrsim10^{10}\,\mathrm{M}_{\odot}$. Results are also very similar in CE-12, where CDM galaxies have lost 80~per~cent of their mass after 9\,Gyr, and SIDM galaxies have lost 90~per~cent.

\begin{figure*}
    \centering
    \includegraphics[width=0.8\textwidth]{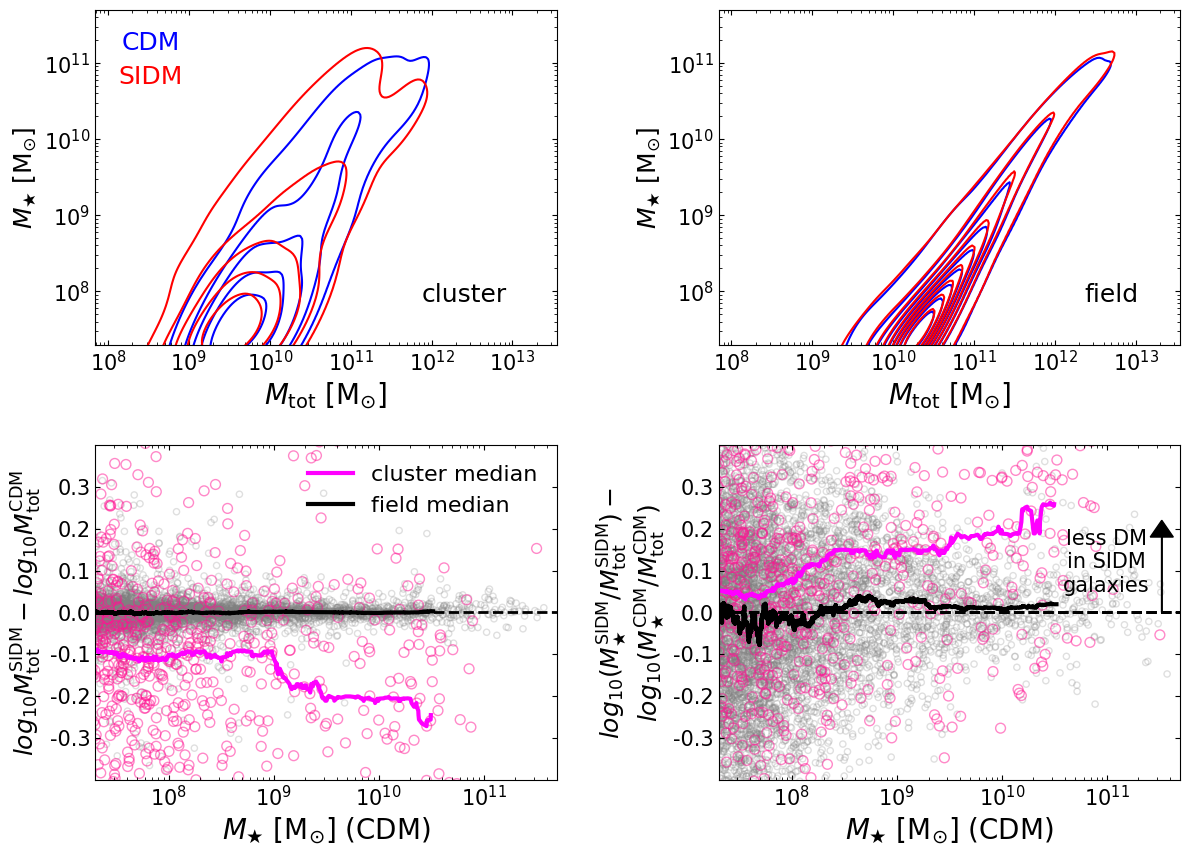}
    \caption{Stellar-to-halo mass relation for galaxy pairs with $>$$10$ star particles ($M_\star\gtrsim5\times10^{7}\,\mathrm{M}_{\odot}$), matched between CDM and SIDM simulations. {\it Top panels}: number-density contours of the stellar mass {\em versus} total mass in cluster CE-12 (left) and the field (right). Both are smoothed with the same circular Gaussian kernel of width $\sigma=0.35$\,dex: the increased scatter inside a cluster is real. A version for all (including unmatched) galaxies looks qualitatively similar.
    {\it Bottom panels}: the difference in total mass (left) and stellar-to-halo mass (right) between the SIDM and CDM galaxy populations. Pink points show matched galaxy pairs in cluster CE-12, with the running median overlaid; grey points show pairs in the field. The effect of SIDM is greatest for more massive galaxies.
    }
    \label{fig:shmr}
\end{figure*}

\section{Observable differences between cluster galaxies in CDM and SIDM}\label{sec:observables}
We saw in Section~\ref{sec:indiv_subs} that a galaxy made of SIDM has a higher rate of DM loss than an identical galaxy made of CDM. However, observations of the real Universe do not have the luxury of matched comparisons to a control sample or null test. In this section we investigate whether the increased rate of mass loss has observable effects on the population of galaxies in a cluster at ${z=0}$.

\subsection{Stellar-to-Halo Mass Relation}\label{sec:fit_SHMR}



\begin{figure*}
    \centering
    \includegraphics[width=0.8\textwidth]{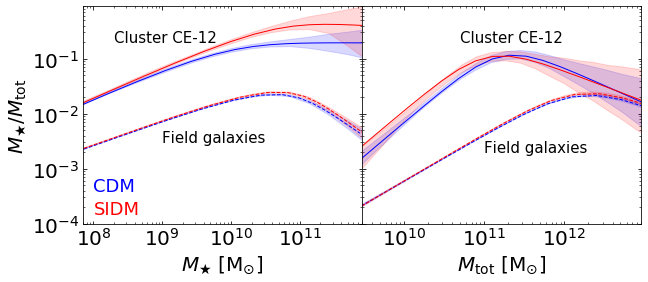}
    \caption{Fits to the SHMRs using equation~\ref{eq:moster}. {\it Left}: the SHMR as a function of stellar mass. Fits to the cluster galaxies in CE-12 are shown as solid lines, and to field galaxies fits as dashed lines. Blue and red lines represent the CDM and SIDM versions of a given simulation respectively. Shaded regions represent the 68~per~cent confidence regions, obtained from the $16^\mathrm{th}$ and $84^\mathrm{th}$ percentiles of the MCMC chain. {\it Right}: similar to the left panel, but now for the SHMR as a function of halo mass. The fits to the galaxies in CE-05 are similar but noisier, because that cluster has fewer member galaxies.}
    \label{fig:shmr_fit}
\end{figure*}

At the mass scale of individual galaxies, the SHMR of field galaxies is indistinguishable between CDM and SIDM simulations (Figs.~\ref{fig:shmr} and \ref{fig:shmr_fit}). This is expected because efficient gas cooling and star formation ensure that a baryon-dominated core retains a deep gravitational well \citep{2019MNRAS.488.3646R}. Once a galaxy falls into a cluster, tidal forces preferentially remove DM, which is more diffuse than stars.

We first investigate the SHMR for matched pairs of galaxies with more than 10 star particles at $z=0$ (Fig.~\ref{fig:shmr}). On average, SIDM cluster galaxies ended up with ${\sim}$0.12\,dex ($25$~per~cent) lower masses than their CDM counterparts. This effect increases to ${\sim}$0.2\,dex ($35$~per~cent) for the most massive cluster member galaxies. We then fit the \citet{2013MNRAS.428.3121M} relation, as described in Section~\ref{sec:SHMR}. We fit all galaxies, not just those matched between simulations (as would be done with observational data). Because this adds some almost-stripped galaxies, this raises the normalisation of the SHMR at low masses by a factor ${\sim}1.5$ for both CDM and SIDM, and moves the location of the turnover within its (considerable) statistical uncertainty. The best fits are shown in Fig.~\ref{fig:shmr_fit}, and the best fitting parameters are listed in Table~\ref{tab:fit_params}. 



The SHMRs for CDM and SIDM cluster galaxies are only distinguishable at the high mass end, when binning by stellar mass. Fortunately, it is possible to measure this observationally. We find that cluster galaxies within 2$R_{200}$ with stellar mass $10^{10-11}\,\mathrm{M}_{\odot}$ have $M_{\star}/M_{\rm{tot}}$ 8~times higher than field galaxies in a CDM universe, but 13~times higher in an SIDM universe. For cluster galaxies within $R_{200}$, we find that these numbers increase to 10 and 20 (the best fitting parameters are included in Table~\ref{tab:fit_params}, but the fits are not shown on Fig.\ref{fig:shmr_fit}). There is considerable scatter in the SMHR at these masses, and it would require noise-free measurements of ${\sim}32$ cluster galaxies to distinguish between these values at 3$\sigma$, assuming that the SHMR for field galaxies is well known. It would be more challenging to measure other quantities like the slope of the SHMR at low masses, or the position of the turnover, because these vary by less than five per~cent with different DM.

\begin{table}
\centering
\caption{The best fit parameters of the SHMR~\ref{eq:moster} for field galaxies and for cluster galaxies within $2R_{200}$ and $R_{200}$ of CE-12. The 68~per cent confidence intervals are the difference between the $16^\mathrm{th}$ and $84^\mathrm{th}$ percentiles of the marginalized 1D posteriors.}
\label{tab:fit_params}
\begin{tabular*}{\columnwidth}{@{}c@{\hspace*{12pt}}c@{\hspace*{12pt}}c@{\hspace*{12pt}}c@{\hspace*{12pt}}c@{}}
\hline
\multicolumn{5}{c}{Field galaxies}\\
\hline
& \multicolumn{2}{c}{Fit to $M_{\mathrm{tot}}$($M_{\star}$)} & \multicolumn{2}{c}{Fit to $M_{\star}$($M_{\mathrm{tot}}$)}\\
& \multicolumn{1}{c}{CDM} & \multicolumn{1}{c}{SIDM} & CDM & SIDM\\
\hline
$\mathrm{log_{10}} M_{1}$ & \multicolumn{1}{|c}{$12.09^{+0.06}_{-0.05}$} & \multicolumn{1}{c|}{$12.11^{+0.06}_{-0.05}$} & $12.22^{+0.05}_{-0.05}$ & $12.2^{+0.05}_{-0.05}$\\
$N$ & \multicolumn{1}{|c}{$0.022^{+0.001}_{-0.001}$} & \multicolumn{1}{c|}{$0.024^{+0.001}_{-0.001}$} & $0.021^{+0.002}_{-0.001}$ & $0.023^{+0.002}_{-0.001}$\\
$\beta$ & \multicolumn{1}{|c}{$0.81^{+0.02}_{-0.02}$} & \multicolumn{1}{c|}{$0.81^{+0.02}_{-0.02}$} & $0.84^{+0.02}_{-0.02}$ & $0.86^{+0.02}_{-0.02}$\\
$\gamma$ & \multicolumn{1}{|c}{$0.46^{+0.04}_{-0.04}$} & \multicolumn{1}{c|}{$0.48^{+0.04}_{-0.04}$} & $0.6^{+0.07}_{-0.07}$ & $0.57^{+0.07}_{-0.07}$\\
$\sigma_{\mathrm{M}}$ & \multicolumn{1}{|c}{$0.215^{+0.002}_{-0.002}$} & \multicolumn{1}{c|}{$0.214^{+0.002}_{-0.002}$} & \hspace{2mm} $0.278^{+0.003}_{-0.003}$ & $0.281^{+0.003}_{-0.003}$\\
\hline\\
\multicolumn{5}{c}{Cluster galaxies ($R<2R_{200}$)}\\
\hline
& \multicolumn{2}{c}{Fit to $M_{\mathrm{tot}}$($M_{\star}$)} & \multicolumn{2}{c}{Fit to $M_{\star}$($M_{\mathrm{tot}}$)}\\
& \multicolumn{1}{c}{CDM} & \multicolumn{1}{c}{SIDM} & CDM & SIDM\\
\hline
$\mathrm{log_{10}} M_{1}$ & \multicolumn{1}{|c}{$10.55^{+0.29}_{-0.24}$} & \multicolumn{1}{c|}{$10.84^{+0.47}_{-0.35}$} & $11.23^{+0.2}_{-0.24}$ & $11.02^{+0.25}_{-0.23}$\\
$N$ & \multicolumn{1}{|c}{$0.1^{+0.02}_{-0.03}$} & \multicolumn{1}{c|}{$0.26^{+0.37}_{-0.11}$} & $0.11^{+0.02}_{-0.02}$ & $0.26^{+0.29}_{-0.03}$\\
$\beta$ & \multicolumn{1}{|c}{$1.27^{+0.13}_{-0.17}$} & \multicolumn{1}{c|}{$1.27^{+0.13}_{-0.22}$} & $1.23^{+0.15}_{-0.18}$ & $1.27^{+0.15}_{-0.18}$\\
$\gamma$ & \multicolumn{1}{|c}{$0.0^{+0.19}_{-0.19}$} & \multicolumn{1}{c|}{$0.08^{+0.42}_{-0.43}$} & $0.66^{+0.34}_{-0.39}$ & $0.08^{+0.37}_{-0.4}$\\
$\sigma_{\mathrm{M}}$ & \multicolumn{1}{|c}{$0.36^{+0.01}_{-0.01}$} & \multicolumn{1}{c|}{$0.41^{+0.01}_{-0.01}$} & $0.62^{+0.02}_{-0.02}$ & $0.41^{+0.03}_{-0.03}$\\
\hline\\
\multicolumn{5}{c}{Cluster galaxies ($R<R_{200}$)}\\
\hline
& \multicolumn{2}{c}{Fit to $M_{\mathrm{tot}}$($M_{\star}$)} & \multicolumn{2}{c}{Fit to $M_{\star}$($M_{\mathrm{tot}}$)}\\
& \multicolumn{1}{c}{CDM} & \multicolumn{1}{c}{SIDM} & CDM & SIDM\\
\hline
$\mathrm{log_{10}} M_{1}$ & \multicolumn{1}{|c}{$10.53^{+0.29}_{-0.25}$} & \multicolumn{1}{c|}{$10.85^{+0.3}_{-0.2}$} & $11.25^{+0.29}_{-0.25}$ & $11.01^{+0.43}_{-0.43}$\\
$N$ & \multicolumn{1}{|c}{$0.14^{+0.04}_{-0.04}$} & \multicolumn{1}{c|}{$0.5^{+0.34}_{-0.14}$} & $0.16^{+0.04}_{-0.05}$ & $0.5^{+0.92}_{-0.09}$\\
$\beta$ & \multicolumn{1}{|c}{$1.28^{+0.14}_{-0.18}$} & \multicolumn{1}{c|}{$1.37^{+0.12}_{-0.17}$} & $1.19^{+0.2}_{-0.24}$ & $1.37^{+0.35}_{-0.55}$\\
$\gamma$ & \multicolumn{1}{|c}{$0.06^{+0.21}_{-0.2}$} & \multicolumn{1}{c|}{$0.41^{+0.38}_{-0.26}$} & $0.6^{+0.44}_{-0.51}$ & $0.41^{+0.53}_{-0.63}$\\
$\sigma_{\mathrm{M}}$ & \multicolumn{1}{|c}{$0.34^{+0.01}_{-0.01}$} & \multicolumn{1}{c|}{$0.39^{+0.01}_{-0.01}$} & $0.62^{+0.03}_{-0.03}$ & $0.39^{+0.04}_{-0.04}$\\
\hline
\end{tabular*}
\end{table}



\subsection{The stripping factor}\label{sec:tau_strip}
Another measure used to express the mass lost from cluster galaxies is the `stripping factor' \citep{2019MNRAS.487..653N}
\begin{equation}\label{eq:tau_strip}
\tau_\mathrm{strip}(M_\mathrm{\star}) \equiv 1 - \frac{\widetilde{M}_\mathrm{tot, cluster}(M_\mathrm{\star})}{\widetilde{M}_\mathrm{tot, field}(M_\mathrm{\star})},
\end{equation}
where $\widetilde{M}_\mathrm{tot, cluster}(M_{\star})$ and $\widetilde{M}_\mathrm{tot, field}(M_{\star})$ are the median total mass of cluster and field galaxies in a bin of stellar mass $M_{\star}$. This definition is motivated by a model in which a galaxy's star formation is quenched as it enters a cluster. Since no new stars are formed, field galaxies of a given stellar mass act as the progenitors of cluster galaxies with the same stellar mass.

\begin{figure}
    \centering
    \includegraphics[width=\linewidth]{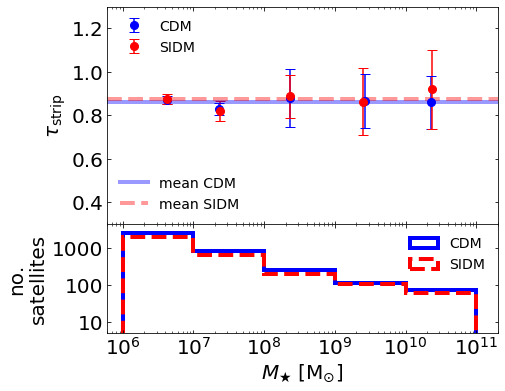}
    \caption{{\it Top}: $\tau_\mathrm{strip}$ (equation~\ref{eq:tau_strip}) as a function of mean stellar mass in five stellar bins. The results for the galaxies in the CDM and SIDM versions of CE-12 are plotted in blue and red respectively. The horizontal solid blue and dashed red line are the mean $\tau_\mathrm{strip}$ of the CDM and SIDM galaxies respectively. The mean stripping factor has a value of $0.86\pm0.03$ and $0.87\pm0.04$ for the CDM and SIDM satellites respectively. The results for cluster CE-05 are $0.83\pm0.04$ and $0.85\pm0.04$. {\it Bottom}: histogram of number of galaxies in the same five stellar bins as plotted in the top panel. Again blue represents CDM and red SIDM. }
    \label{fig:tau_strip}
\end{figure}

We split our sample of cluster (CE-12) and field galaxies into logarithmic bins of 1\,dex in stellar mass ranging from $10^{6}$ to $10^{11}\,\mathrm{M}_{\odot}$, and calculate the stripping factor in each bin; the result is shown in Fig.~\ref{fig:tau_strip}. The errors on the stripping factors are calculated using bootstrapping.


The difference between CDM and SIDM is not significant in this measure, although the largest hint of a difference again appears to be in galaxies with high stellar mass. 
The mean stripping factor of galaxies inside $2R_{200}$ at $z=0$ is $0.86\pm0.03$ and $0.87\pm0.04$ for the CDM and SIDM version of cluster CE-12 respectively (blue solid and red dashed horizontal lines in Fig.~\ref{fig:tau_strip}), and there is little scatter about this value in the different stellar mass bins. For massive galaxies with $10^{10}M_{\sun}<M_{\star}<10^{11}M_{\sun}$, the mean stripping factor for SIDM is $\mathcal{O}(10^{-2}$) higher than for CDM, but this is much smaller than statistical uncertainty. More stripping occurs in the inner parts of the cluster, and the stripping factors rises to $0.88\pm0.03$ and $0.90\pm0.05$ for galaxies inside $R_{200}$. Again there is little hope for observational discrimination.

Stripping factors are reduced in the lower mass cluster CE-05, to $0.83\pm0.04$ and $0.85\pm0.04$ for the CDM and SIDM versions of galaxies within 2$R_{200}$ with again little scatter about these values. A more massive cluster seems to increase slightly both the stripping of mass and the effect of self-interactions. 

\subsection{The number and radial distribution of cluster galaxies}

\begin{figure}
    \centering
    \includegraphics[width=0.8\linewidth]{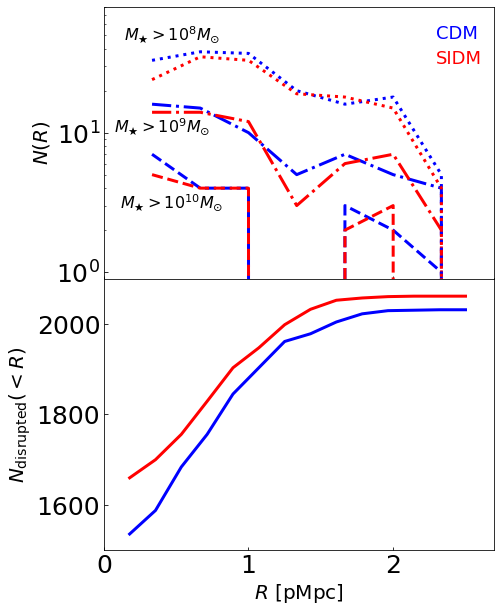}
    \caption{\textit{Top}: The radial distribution of galaxies that survive until $z=0$, in CDM and SIDM versions of cluster CE-05. The only useful difference is the slight reduction of SIDM galaxies inside the cluster core. \textit{Bottom}: The last known location of galaxies that did not survive until $z=0$. Cumulative number of galaxies inside a given radius, in the simulation snapshot immediately before they were disrupted.}
    \label{fig:disrupt_hist}
\end{figure}

There are ${\sim}20$~per~cent fewer member galaxies in the SIDM version of a given cluster at ${z=0}$ (Table~\ref{tab:clust_details}). Most of the discrepancy is in the central $\sim100$\,kpc, which is also where the most disruption takes place of SIDM galaxies whose CDM counterparts survive (Figure~\ref{fig:disrupt_hist}). This is consistent with our earlier findings that SIDM barely changes the orbits of galaxies, but makes them more susceptible to disruption (Section~\ref{sec:pop_subs}). Cluster outskirts contain similar numbers of galaxies, with the populations continually replenished by objects infalling from the field.

It would be difficult to distinguish between CDM and SIDM using cluster richness, given the intrinsic scatter in the mass-richness relation \citep{2017MNRAS.466.3103S,2019PASJ...71..107M,2021ApJS..253....3H}. It is probably also difficult to distinguish between CDM and SIDM using the radial distribution of cluster galaxies. We find that 
33~per~cent and 36~per~cent of galaxies reside inside $0.5R_{200}$ in the CDM version of clusters CE-05 and CE-12, compared to 30~per~cent and 26~per~cent in the SIDM versions. More simulations are needed to determine the population mean and intrinsic scatter, but the difference is likely to be washed out by projection effects (of outlying members in front of/behind the cluster core, and field galaxies onto cluster outskirts).

\section{Discussion \& Conclusions}\label{sec:conclusions}
We studied the effects of self-interactions on the mass stripping of
galaxies as they fall into galaxy clusters by comparing cosmological
simulations with and without DM self-interactions. When a galaxy falls
into a cluster, DM interactions accelerate the rate of mass
stripping. Over 33~per~cent of galaxies in an SIDM cluster can be
entirely disrupted by the present time, compared to 20~per~cent in a
CDM cluster. Unfortunately, the disrupted galaxies (which are the most
different between CDM and SIDM) are no longer observable. The orbits
of surviving galaxies are essentially unchanged, and disrupted
galaxies are continually replaced by new ones falling into the
cluster. When comparing matched galaxies between the CDM and SIDM
versions of a given cluster (Section~\ref{sec:indiv_subs}), we find
significant differences in mass loss. However, when we only look at
the population of galaxies remaining in the cluster at ${z=0}$
(Section~\ref{sec:observables}), we find considerably smaller
differences. SIDM galaxies are more susceptible to disruption, so
there is a large group of disrupted SIDM galaxies which does not
contribute to the signal at ${z=0}$. 

Potentially observable ways to discriminate between CDM and SIDM
include the (high mass normalisation of the) stellar-to-halo mass
relation of galaxies in clusters, compared to galaxies in the field,
or the stripping factor, both of which describe the mass of the DM in
a galaxy of fixed stellar
mass. 
We found a 25~per~cent increase in the ratio of stellar-to-total mass
of SIDM galaxies with stellar mass
$M_\star>5\times10^{7}\,\mathrm{M}_{\odot}$. 
The absolute normalisation of the relation is likely to be needed to
discriminate SIDM from CDM, but this depends to some extent on the
subgrid physics of the simulations. However, as in the field the
relation is nearly indistinguishable for a CDM and SIDM universe, one
could use the difference between the field and cluster relations at a
given stellar mass to try and discriminate between the two
models. From the left panel of Fig.~\ref{fig:shmr_fit}, we find that,
at approximately the stellar mass of the Milky Way,
$10^{10.5}\,\mathrm{M}_{\odot}$, the ratio
$M_{\star}/M_{\mathrm{tot}}$ is 8 and 13 times higher in the cluster
compared to the field for the CDM and SIDM versions of CE-12
respectively.

Previous, 
DM-only simulations \citep{2021arXiv210608292B} 
predicted larger differences between SIDM and CDM, probably because of the way stars were assigned to galaxies after the simulation using a semi-analytic model. 
In DM-only SIDM simulations subhaloes form cores more easily than when baryons are included, making them more easily disrupted. 
In contrast, our simulations co-evolved a population of baryons and SIDM.
In the full hydrodynamical simulation a large number of cluster galaxies fail to form cores or have their cores re-contracted by baryons, and so they are more durable.



In the future, it would be informative to simulate more SIDM clusters.
While the {\sc c-eagle} suite comprises 30 simulated CDM clusters, only
two have been re-run with
SIDM. 
It is also important to note that a cross-section of $\sigma/m = 1$
$\mathrm{cm}^{2}\mathrm{g}^{-1}$ has arguably already been ruled out
at the $\mathcal{O}(1000\,\mathrm{km}\,\mathrm{s}^{-1}$) collision
velocities between particles typical in clusters. Performing the same
tests with simulations for  a lower cross section would presumably
produce smaller differences and would require even higher
signal-to-noise observations. Future surveys, such as the Euclid \citep{laureijs2011euclid}, Rubin Legacy Survey of Space and Time \citep[LSST;][]{2009arXiv0912.0201L}, SuperBIT \citep{2018SPIE10702E..0RR}, and JWST Cosmos-Webb survey (C.\ Casey \& J.\ Kartaltepe, pers.\ comm.\ 2021)
will provide data with higher signal-to-noise than ever before,
potentially making such tests possible.

\section*{Acknowledgements}

We thank Anna Niemiec for discussing the physics of infall, and Yannick Bah\'e and David Barnes for making the {\sc c-eagle} data available.

ES and RM are supported by the Royal Society. AR is supported by the European Research Council’s Horizon2020 project `EWC' (award AMD-776247-6). KAO and CSF acknowledge support by the European Research Council (ERC) through Advanced Investigator grant to CSF, DMIDAS (GA 786910). 
This work used the DiRAC@Durham facility managed by the Institute for Computational Cosmology on behalf of the STFC DiRAC HPC Facility (www.dirac.ac.uk). The equipment was funded by BEIS capital funding via STFC capital grants ST/K00042X/1, ST/P002293/1, ST/R002371/1 and ST/S002502/1, Durham University and STFC operations grant ST/R000832/1. DiRAC is part of the National e-Infrastructure.


\section*{Data Availability}
The CDM simulations were introduced by \cite{2017MNRAS.470.4186B}; the SIDM simulations by \cite{2018MNRAS.476L..20R}. Data from those works are available from the original papers.

\bibliographystyle{mnras}
\bibliography{references}

\appendix

\bsp	
\label{lastpage}
\end{document}